# Super-resolution-enabled atmospheric tomography for astronomical multi-wavefront-sensor adaptive-optics systems

**Carlos M. Correia**[1,2] · **Pierre Jouve**[3,4] · **Jesse Cranney**[5] · **Guido Agapito**[6] · **Cédric Taïssir Heritier**[7]

© The Author(s) 2025

## Abstract

Recent work by Oberti et al, (*Astron. Astrophys.*, *667*, 48, 2022) argued and made a compelling case that classical astronomical adaptive optics (AO) tomography performance can be further enhanced by carefully designing and optically configuring the system to leverage inherent super-resolution (SR) capabilities. Our goal here is to further materialise the concept by providing the means to compute SR-enabling tomographic reconstructors for AO and showcase its broad uptake on soon every 10 m-class VIS/NIR telescopes and Giant Segmented Mirror Telescopes of up to 40 m in diameter. To that end we indicate the necessary tomography generalisations where we: *(i)* clarify how *model-and-deploy* is a generic methodological umbrella for linear minimum-mean-squared-error (LMMSE) tomographic reconstructors arising naturally from the solution of the tomographic inverse problem, thus unifying various solutions presented as distinct in the literature within a single framework, *(ii)* recall how such solutions are found as limiting cases of a model-based optimal control problem, thus elucidating how *pseudo-open-loop control* is a feature of the latter that allows LMMSE reconstructors to be adapted to closed-loop systems, *(iii)* review the two forms of the LMMSE tomographic reconstructors, highlighting the necessary adaptations to accommodate super-resolution, *(iv)* review the implementation in either dense-format vector-matrix-multiplication or sparse iterative forms and *(v)* discuss the implications for runtime and off-line real-time implementations, anticipating widespread adoption. We illustrate our examples with physical-optics numerical simulations for 10 m and 40 m-scale systems showing the performance benefits of super-resolution in the order of several tens of nm rms and the computational burden associated.

---

Extended author information available on the last page of the article

 Springer

# 1 Introduction to model & deploy tomography

## 1.1 Atmospheric tomography: The global context

After the principle leading to the increase of the isoplanatic patch using multi-conjugation was first formulated by [9] it became clear that the underlying problem was one of tomography [72] whereby the three-dimensional structure of the atmosphere needed be estimated prior to projection onto a single or multiple deformable mirrors optically conjugated at different ranges.

The core tomography problem in AO deviates considerably from classical tomography – here the practical method of estimating a three-dimensional object from many line integrals over the circle, building on the famous Radon transform, named after the seminal work in integral geometry by [63], in at least four key points:

1. limited-angle (typically less than 2 arcmin) resulting in extremely ill-posedness [30],
2. small number of projections (commonly 4-8), whereas hundreds are typically advocated for reaching robust solutions [43],
3. agency upon the object via adaptive optics deformations,
4. real-time operation to keep up with the ever evolving atmospheric turbulence.

Linearity of measurements, stability, robustness and overall system complexity led to the adoption of negative-feedback control systems turning the problem into one of dynamic wavefront estimation and (optimal) control rather than one of static turbulence estimation (see Fig. 1).

The AO problem, with its intrinsic features, lies somewhat between the two knowledge domains typically employed to solve it in a efficient and practical manner. These can loosely be classified as

> *(i)* computerised tomography / inverse problem approaches, dealing only with the wavefront reconstruction
> *(ii)* stochastic optimal control, where the wavefront reconstruction and control are approached *jointly*

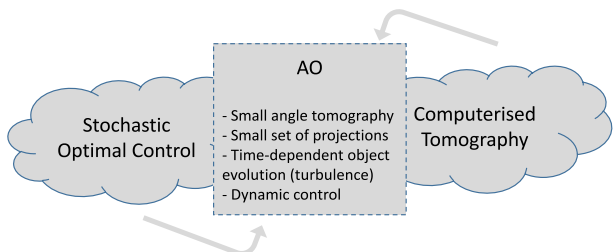

**Fig. 1** Conceptual sketch of background frameworks used to tackle the fundamental tomography problem at the core of wide-field AO

Seminal works under *(i)* formulate the problem as one of (image) reconstruction from its projections by minimising the residual wavefront variance – which is effectively equivalent to maximising the Strehl-ratio [42]. Most prominent examples are [28, 32] who provide a linear minimum-mean-squared-error (LMMSE) solution expressed in different spaces, modal Zernike polynomials the former and zonal or bilinear splines the latter. Gavel et al [35] adopted alternatively the Fourier basis and was the first to establish a link to a well-known back-projection solution in classical tomography. The actual implementation was then further extended to exploit sparsity and matrix structure to cope with the massive numbers of degrees-of-freedom on Giant Segmented Mirror Telescopes: see [30] for an earlier review and [68] for a more recent one where the reconstruction is recast within the more classic algebraic reconstruction technique (ART), but also [54, 73], all exploiting the (sparse) matrix structure in a given descriptor space. Such efforts were adapted to closed-loop using the concept of pseudo-open-loop where the full atmospheric profile estimation is recast in real-time by adding to the (noisy) residual wavefront measurements the previously-computed deformable mirror correction in an appropriate control space. The whole scheme is made stable by careful choice of time filtering coefficients [38].

On the other hand, in *(ii)* the problem was approached as one of optimal stochastic control using a Linear-Quadratic-Gaussian (LQG) formalism [46] (the LQG had been suggested before for AO control, but not for the tomographic case though) with several subsequent improvements and lab demonstrations [22, 59].

It is to be noted that the formal solution in *(i)* can be obtained as a limiting case of *(ii)* [21, 60]. The optimal stochastic control solution can thus be seen as a more general framework where tomography is considered within the broader scope of linear, negative-feedback, dynamic control systems. Despite this opening to a potential performance leap, the caveat often attributed to such solutions is the added complexity for real-time implementation and the model sizes which tend to be larger than in the static formulation.

It is therefore no surprise that several contributions made the effort of bridging the gap by making static reconstruction solutions in *(i)* more adapted to the underlying dynamic control problem [24, 61]; in parallel, the opposite path was also pursued, *i.e.* adapting the optimal control solution to large number of degrees-of-freedom, real-time tomographic problems [18, 39, 49].

## 1.2 Model&Deploy tomography

The framework we place ourselves in is one of *Model&Deploy* which consists in first compiling the reconstructor model either from identification, fit or analytics in soft-real-time, i.e. hundreds to thousands of times slower than its deployment at runtime speeds, often in the hundreds to thousands of Hertz.

In other words, *Model&Deploy* refers to solving the tomographic problem in two stages. First a *sufficiently accurate* forward model is formulated, where *sufficiently accurate* is loosely defined as a model capable of adhering to predefined criteria that ensure suitable optical performance and efficient numerical processing, retaining the main features of the real problem, typically linearity, sparsity, block-structure etc., and, subsequently, inverting the said model *off-line* using appropriate methods (e.g.

singular-value decomposition, regularisation, etc), Alternative iterative implementations have been the subject of great attention and we shall cover them also further below.

On a second stage, the reconstructor resulting from the inversion is applied in real-time to either a closed-loop (general case) or open-loop measurements in special AO correction cases.

We aim to clearly establish that this is a generic framework applicable not only to our problem but also to other contexts. Many approaches in atmospheric tomography, despite being referred to by different names, are not fundamentally distinct solutions. Instead, they are all specific instances of the broader *Model & Deploy* framework. Approaches like *Learn & Apply* are not exceptional; rather, they are particular cases where models are adjusted to a working space (e.g., measured slopes) using model-fitting functions on empirical data acquired during observations. Other examples include Fourier-based reconstructors [35], modal reconstructors [17], and various others cited earlier in the text, which often leverage an intermediate estimation space for computational convenience.

### 1.3 Super-resolution wavefront reconstruction

Recent work in [55] argued that *classical* astronomical adaptive-optics tomography performance can be further improved by specifically configuring the optical system to add to and exploit any built-in super-resolution (SR) capabilities. An early illustration was provided by [31] and adoption is becoming widespread.

SR refers to both sampling and reconstruction techniques whereby the tomographic system can probe & reconstruct spatial-frequencies beyond the nominal Nyquist-Shannon cut-off imposed by the individual sensor sampling step size (Shack-Hartmann sub-aperture in the AO applications covered here).

We note that SR is natively granted in multi-WFS tomographic systems, both **statistically** using regularised reconstruction in the form of an LMMSE or a MAP (Maximum-a-Posteriori) estimator and **geometrically** as the projections of lenslet sub-apertures along different lines-of-sight generate a height-dependent sampling offset akin to a linear scanning, thus improving through proper reconstruction the effective Nyquist-Shannon frequency beyond that of the individual sensors when considered isolated. This has been exploited by [76] without expressing it as a super-resolution capability but over-sampling instead, therefore not exploiting it to its fullest.

We suggest that the built-in SR capabilities of a tomographic AO system can be further engineered by breaking geometric symmetries and regularities, allowing for increased measurement diversity. This is done in two steps: 1) optimising the sampling method (e.g. rotating and offsetting the sensors, using non-homogeneous sampling, using uneven asterisms etc – something we already covered in [55]) and 2) generalising the wave-front reconstructors to cope with such modifications, without which no signal recovery could be possible.

## 1.4 Goals of this paper

This paper reports on how to instantiate the SR-enabling tomographic reconstructors using a *zonal* representation. We have two goals in mind: *(a)* provide detail about the synthesis of the tomographic reconstructors and *(b)* provide illustrative examples with more instances of 10 m-class and 20-40 m-class systems.

In summary, this paper:

- Provides an overview of the optimal, minimum-variance control solution and how the static MMSE + POLC is derived as a limiting solution under certain assumptions
- Reviews the two forms of the MMSE and recalls the discrete (zonal) formulations, highlighting the necessary adaptations to accommodate super-resolution
- Reviews the implementation in either dense vector-matrix multiply (VMM) or iterative forms
- Provides sample numerical simulations for 10m and GSMTs showing the performance benefits of super-resolution

## 1.5 Paper organisation

This paper is organised as follows. In section §2 we start from the generic state-space, model-based, optimal control solution to find limiting cases and recall the two LMMSE formulations often adopted by adaptive optics practitioners. In section §3 we develop super-resolution capable formulations and show that the classical formulations need only be little modified to cope with SR. In section §4 we discuss iterative and dense matrix implementations and in section §5 we give performance results for 8-10 class telescopes. We conclude with illustrations for the systems Harmoni and Morfeo for the ELT in section §6.

# 2 Atmospheric tomography: Wavefront reconstruction & control in closed-loop

We start by outlining a general solution derived from optimal stochastic control theory to show how its simplification leads to well-established and practical formulations.

## 2.1 Strehl-optimal cost-functional

We begin by recalling the goal of tomography, which is to provide the best estimation of a 3D object (in our case the stratified atmospheric wave-front phase profiles) from its noisy measurements (projections) along multiple lines-of-sight.

In the case of atmospheric tomography, the cost-functional is expressed as the minimum averaged, pupil-integrated, piston-removed residual phase error $\langle \sigma^2 \rangle_{t,\beta}$ where $\langle \cdot \rangle_{t,\beta}$ represents ensemble averaging over wavefront and noise statistics, con-sidered across time $t$ and covering a field-of-view of interest, discretised with multiple directions $\beta_i$ of which one is usually on-axis.

This choice of cost-functional serves as an indirect measure of optimal image quality. According to the Maréchal approximation, the Strehl-ratio is a decreasing function of the residual phase variance (in rad$^2$), $S = e^{-\langle \sigma^2 \rangle}$ [67].

In mathematical terms, the Strehl-optimal reconstructor, minimises

$$\mathrm{R}^* = \arg \min_{\mathrm{R}} \left\langle \sigma^2 + \kappa \|\mathrm{u}\|^2 \right\rangle_{t,\boldsymbol{\beta}}, \tag{1}$$

where a regularisation term $\kappa \|\mathrm{u}\|^2$ was added to avoid numerical instability when solving for the problem in case singularities in the deformable mirror command subspace arise.

Further expanding (1), the criterion is evaluated as the Euclidean norm $L_2$ over the telescope pupil $\Omega$ of the difference between the input wave-front $\phi_{\beta_i}$ and the deformable-mirror (DM)-produced correction $\phi_{\beta_i}^{dm}$. The aperture-plane residual wave-front error in one single optimisation direction $i$ is

$$\sigma_{\beta_i}^2 = \left\| \boldsymbol{\phi}_{\beta_i} - \boldsymbol{\phi}_{\beta_i}^{dm} \right\|_{L_2(\Omega)}^2, \tag{2}$$

whereas $\sigma^2$ in (1) is evaluated by discretising and weighted averaging over a given field-of-view of interest (contained in the vector $\boldsymbol{\beta}$). In (2) the directional phase $\phi_{\beta_i}$ is obtained by tracing rays (represented by $\mathrm{T}$) through the atmospheric wave-front profiles to the aperture-plane phase grid intercepts in directions $\beta_i$ by interpolating and summing the layered phase distortion in the volume, *i.e.*

$$\boldsymbol{\phi}_{\beta_i} = \sum_{l=1}^{N_l} \mathrm{T}_{\beta_i,l} \boldsymbol{\varphi}_l, \tag{3}$$

or in compact matrix format $\phi_{\beta_i} = \mathrm{T}_{\beta_i} \boldsymbol{\varphi}$ with $\boldsymbol{\varphi}$ a vector of stratified turbulent coefficients stacked together from multiple altitudes.

## 2.2 Minimum-variance solution using stochastic control theory

The optimal stochastic control framework relies (as often found in physics and engineering problems) on establishing and populating a suitable differential model of the underlying physical process(es). We will adopt a discrete-time, state space representation (effectively a difference equation)

$$\begin{cases} \mathrm{x}_{k+1} &=& \mathcal{A}\mathrm{x}_k + \mathcal{B}\mathrm{u}_k + \mathcal{V}\boldsymbol{\nu}_k \\ \mathrm{s}_k &=& \mathcal{C}\,\mathrm{x}_k + \mathcal{D}\mathrm{u}_{k-d} + \boldsymbol{\eta}_k \end{cases}, \tag{4}$$

where $x_k$ is the state vector that contains the wave-front coefficients (or any linearly related variables), $s_k$ are the noisy measurements provided by the WFS in the guide-star (GS) direction; $\nu_k$ and $\eta_k$ are spectrally white, Gaussian-distributed state excitation and measurement noise respectively. In the following we denote cross covariance matrices between $n$ and $m$ as $\Sigma_{mn}$ and assume that $\Sigma_{\nu\nu} = \langle \nu\nu^{\mathrm{T}} \rangle$, $\Sigma_{\eta\eta} = \langle \eta\eta^{\mathrm{T}} \rangle$ are known and that $\Sigma_{\eta\nu} = \langle \eta\nu^{\mathrm{T}} \rangle = 0$. The definition of the model parameters $\{\mathcal{A}, \mathcal{B}, \mathcal{C}, \mathcal{D}\}$ is obtained from physical modelling of the system dynamics and response functions [23].

### 2.2.1 Separation theorem and optimal estimation

The separation theorem between estimation and correction (control) states that the optimal control actions are computed from the conditional mean estimate of the state (here, loosely speaking, the atmospheric wave-front or a linear function thereof). It is readily provided by a Kalman filter, when assuming linearity, a quadratic criterion and Gaussian distributed disturbances (although the latter is not strictly mandatory), as

$$u_k = -\mathcal{K}^{\mathrm{opt}} \widehat{x}_{k|k-1}, \tag{5}$$

with $\widehat{x}_{k|k-1}$ the conditional mean of the state $x_k$ at instant $'k'$ knowing all information up to $'k-1'$. Several works such as [23, 45, 64] provide further technical details on the assumptions needed and the steps followed to reach this solution.

For our discussion here, we shorten the presentation and state the well-established solution to recursively compute the state as

$$\widehat{x}_{k|k-1} = \mathcal{A}\widehat{x}_{k-1|k-2} + \mathcal{B}u_{k-1} + \mathcal{A}\mathcal{H}^{\mathrm{opt}} \left( s_{k-1} - \widehat{s}_{k-1|k-2} \right), \tag{6}$$

where the gain

$$\mathcal{H}^{\mathrm{opt}} = \Sigma \mathcal{C}^{\mathrm{T}} \left( \mathcal{C}\Sigma\mathcal{C}^{\mathrm{T}} + \Sigma_\eta \right)^{-1}, \tag{7}$$

is obtained by solving the so-called estimation Riccati algebraic equation, from which we compute a steady-state estimation error covariance matrix

$$\Sigma = \mathcal{A}\Sigma\mathcal{A}^{\mathrm{T}} + \Sigma_\nu - \mathcal{A}\Sigma\mathcal{C}^{\mathrm{T}} \left( \mathcal{C}\Sigma\mathcal{C}^{\mathrm{T}} + \Sigma_\eta \right)^{-1} \mathcal{C}\Sigma\mathcal{A}^{\mathrm{T}}. \tag{8}$$

For the general case with mirror dynamics (i.e. "slow" with respect to the AO loop bandwidth), estimating $u_k$ via $\mathcal{K}^{\mathrm{opt}}$ and $\widehat{x}_{k|k-1}$ involves the matrix solutions of discrete-time *control* in addition to the *estimation* Riccati equation (shown in (8)) that we shall not develop beyond this point for we are interested in focusing on the core tomographic problem. Such general solutions falling under the "minimum-variance" stochastic optimal control umbrella can be found in [23, 39, 47, 64] and certainly many more. An excellent textbook where the general framework can be found is [69].

### 2.2.2 Pseudo-open-loop control

An intrinsic property of this solution that is seldom brought to light is the actual computation of a pseudo-open loop measurement vector, despite the closed-loop operation. This can be seen by rearranging terms in (6) and expanding $\widehat{s}_{k-1|k-2}$ with the use of the bottom equation in (4) and noting the appearance of a term $s_k - \mathcal{B}u_k$ which is the underlying variable in the pseudo-open-loop control approach of [38]. As we shall further see next, limiting solutions can readily be found from the optimal control solution.

### 2.2.3 Limiting case: Static, linear minimum-mean-square-error solution

It can be shown that under certain assumptions, such as $\mathcal{A} = 0$ in (8), no intrinsic loop delays, infinitely fast DM dynamics and open-loop regulation, leads to $\Sigma = \Sigma_{\boldsymbol{\varphi}}$, after which $\mathcal{H}^{\mathrm{opt}}$ boils down to the linear minimum-mean-squared error (LMMSE) estimator and $\mathcal{K}^{\mathrm{opt}}$ to a deterministic least-squares projection of the estimated wavefronts onto the span of the DM influence functions [64]. This sub-optimal solution is the expression found for the tomographic reconstructor when approached as a static, open-loop inverse problem, and one that leads to an exceedingly simple formulation as follows.

Let the generic system of equations

$$\begin{cases} s = & Mx + \boldsymbol{\eta} \\ x = & Nu \end{cases}. \tag{9}$$

The LMMSE solution writes [1]

$$\begin{aligned} u &= FWs \\ &= N^{\dagger} \boldsymbol{\Sigma}_{xs} \left( \boldsymbol{\Sigma}_{ss} + \boldsymbol{\Sigma}_{\eta\eta} \right)^{-1} s \end{aligned} \tag{10}$$

which is a two-step solution involving a deterministic F *fitting matrix* to DM actuator space which is independent from the statistics of the WFS measurement noise and the atmospheric turbulence, and W the *wave-front tomographic reconstructor*, the product of a cross-covariance matrix $\boldsymbol{\Sigma}_{xs}$ by the inverse of the measurements auto-covariance matrix $(\boldsymbol{\Sigma}_{ss} + \boldsymbol{\Sigma}_{\eta\eta})$. Ways in which the determination of these matrices have been approached for atmospheric tomography are detailed next.

### 2.3 Stochastic wave-front estimation

We now turn our attention to deriving the LMMSE solution for atmospheric tomography. We will distinguish two tomographic cases which solve for the same functional yet differ in their output quantities of interest:

1. *Implicit reconstruction*, also known as *spatio-angular reconstructor* [17, 18] or *Learn&Apply* [74]: Bi-dimensional *aperture-plane* reconstruction which admits a compact representation for it only estimates wavefronts at the pupil plane and never in the stratified volume. This formulation follows directly from (9)

2. *Explicit reconstruction* [29, 32, 66]: Classical, three-dimensional *sliced-volume* reconstruction which estimates wave-fronts across the stratified volume. As we shall see, it uses a modified, albeit mathematically equivalent formulation to (9)

For historical reasons, we shall start with the *Explicit reconstruction* first. Not only it was developed earlier in the late 90's and 2000', it is conceptually closer to the principle of tomography, which is to estimate a three-dimensional, discretised object (in our case, the wavefront profiles).

### 2.3.1 Explicit three-dimensional Wavefront reconstruction LMMSE formulation

In a perfectly linear setting, a stratified wave-front profile $\boldsymbol{\varphi}$ is estimated from its measured projections

$$\widehat{\boldsymbol{\varphi}} = \mathrm{W}\mathrm{s}_{\boldsymbol{\alpha}}, \tag{11}$$

where a hat symbol is used to denote estimated variables, here $\widehat{\boldsymbol{\varphi}}$, and we dropped the time indexing for clarity. The LMMSE estimator W observes $\partial\sigma^2/\partial\mathrm{W} = 0$ and $\mathrm{s}_{\boldsymbol{\alpha}}$ are vector concatenations of linear, noisy measurements from $N_{gs}$ WFSs observing as many GSs along directions $\boldsymbol{\alpha}$ (typically distinct from $\boldsymbol{\beta}$). The measurements are assumed to follow the linear relationship

$$\mathrm{s}_{\boldsymbol{\alpha}} = \mathrm{G}\mathrm{T}_{\boldsymbol{\alpha}}\boldsymbol{\varphi} + \boldsymbol{\eta}, \tag{12}$$

where G is a WFS linear model. For the often used Shack-Hartmann WFS it represents the gradient operation; $\mathrm{T}_{\boldsymbol{\alpha}}$ is a ray-tracing from layers to the intercepts on the telescope pupil. This term often takes the form of a projector in modal tomography, yet its purpose remains the exact same under a zonal formulation; in passing, we note any offsets and rotations are conveniently factored into this term as we shall demonstrate later; $\boldsymbol{\eta}$ represents measurement noise.

The solution is a straight application of the LMMSE formulation (the subscript "L" represents the explicit Layer estimation)

$$\begin{aligned} \mathrm{W}_{\mathrm{L}} &= \boldsymbol{\Sigma}_{\varphi\varphi}\mathrm{T}_{\boldsymbol{\alpha}}^{\mathrm{T}}\mathrm{G}^{\mathrm{T}}\left(\mathrm{G}\mathrm{T}_{\boldsymbol{\alpha}}\boldsymbol{\Sigma}_{\varphi\varphi}\mathrm{T}_{\boldsymbol{\alpha}}^{\mathrm{T}}\mathrm{G}^{\mathrm{T}} + \boldsymbol{\Sigma}_{\eta\eta}\right)^{-1} \\ &= (\mathrm{T}_{\boldsymbol{\alpha}}^{\mathrm{T}}\mathrm{G}^{\mathrm{T}}\boldsymbol{\Sigma}_{\eta\eta}^{-1}\mathrm{G}\mathrm{T}_{\boldsymbol{\alpha}} + \boldsymbol{\Sigma}_{\varphi\varphi}^{-1})^{-1}\mathrm{T}_{\boldsymbol{\alpha}}^{\mathrm{T}}\mathrm{G}^{\mathrm{T}}\boldsymbol{\Sigma}_{\eta\eta}^{-1} \end{aligned} \tag{13}$$

where $\mathrm{s}_{\boldsymbol{\alpha}}$ has been replaced by its expression in (12) and $\boldsymbol{\Sigma}_{\varphi\varphi} = \langle\varphi\varphi^T\rangle$ is a covariance matrix of the layered phase distortion in the volume computed by evaluating (23) over all possible baseline combinations of estimated samples / or between modal coefficients when a modal approach is employed as in [32]. Phase disturbances in each layer is assumed to be independent from each others, hence $\boldsymbol{\Sigma}_{\varphi\varphi}$ and $\boldsymbol{\Sigma}_{\varphi\varphi}^{-1}$ are block diagonal matrices.

In parallel the same formulation was developed for NFIRAOS, the Thirty Meter Telescope's multi-conjugate Narrow Field InFrared Adaptive Optics System [77]. Although equivalent in principle, [28] made groundbreaking developments when suggesting the adoption of a zonal approach and the formula in (12) which admits a sparse representation, in an attempt to solve the problem iteratively to circumvent the untenable computational complexity at the time of its development. Subsequent hardware advancements have significantly reduced the complexity of real-time deployment, diminishing the necessity for such implementations. A detailed discussion of these advancements is provided in Section 4.

Further studies were conducted to jointly tackle the computational complexity and the lack of optimality in closed-loop, for example using distributed minimum-variance control [39]. We note that the tomography reconstructor adopted for NFIRAOS uses a so-called over-sampling factor of 2, which effectively grants a wave-front reconstructed at a higher resolution, therefore exploiting the inherent SR capabilities of a multi-sensor system. However, to our understanding this was done without referring to the SR concept which could have opened the way to control its DMs with a lower density WFS for instance. This option was neither exerted nor studied.

### 2.3.2 Implicit, compact, spatio-angular LMMSE reconstruction

For ground-conjugated systems (GLAO, LTAO and MOAO) – but, as we shall see, not limited to – an alternative implicit formulation can be adopted where the layered wavefront profiles are never explicitly estimated, only their pupil-plane counter-parts along lines-of-sight of interest.

Using the definition of the LMMSE in (10) and defining the pupil-plane phase along direction $\boldsymbol{\alpha}$ as $\phi(\boldsymbol{\alpha}) = \mathrm{T}_{\boldsymbol{\alpha}}\varphi$, we obtain the so-called *spatio-angular* (SA) reconstructor

$$
\begin{aligned}
\mathrm{W}_{\mathrm{SA}} &= \overline{\boldsymbol{\Sigma}}_{\phi s} \left(\boldsymbol{\Sigma}_{ss} + \boldsymbol{\Sigma}_{\eta\eta}\right)^{-1} \\
&= \overline{\boldsymbol{\Sigma}_{\phi_{\boldsymbol{\beta}}\phi_{\boldsymbol{\alpha}}}\mathrm{G}^{\mathrm{T}}} \left(\mathrm{G}\boldsymbol{\Sigma}_{\phi_{\boldsymbol{\alpha}}\phi_{\boldsymbol{\alpha}}}\mathrm{G}^{\mathrm{T}} + \boldsymbol{\Sigma}_{\eta\eta}\right)^{-1} \\
&= \overline{\mathrm{T}_{\boldsymbol{\beta}}}\mathrm{W}_{\mathrm{L}}
\end{aligned}
\tag{14}
$$

where we denote $\boldsymbol{\Sigma}$ auto- and cross-covariance matrices of the vectors indicated in subscript, and $\overline{\boldsymbol{\Sigma}}$ an averaged version over many directions and $\overline{\mathrm{T}_{\boldsymbol{\beta}}} = \sum_{i=1}^{n_{opt}} w_i \mathrm{T}_{\boldsymbol{\beta}_i}$ over $n_{opt}$ optimisation direction $\boldsymbol{\beta}$.

The simplification entailed by (14) – which, in hindsight, is actually a more straightforward formulation stemming directly from (10) – consists in estimating first a weighted aperture-plane wave-front profile followed by the fitting step – a compact form which we dub the **implicit tomographic reconstructor**.

Following the properties of the LMMSE discussed under §2.2.3, this estimation step is independent from the DM fitting (number, conjugation and geometry).

The advantages of this formulation for real-time processing are the following:

1. The reconstructor implicitly relies on a richer model estimating a large number of layers without ever making their estimation explicit.
2. The latent space is typically of a smaller size leading to a more convenient memory footprint in the soft-real-time computer

3. The slopes' covariance matrix $\Sigma_{ss}$ as part of the reconstructor, can be estimated directly from on-sky measurements (in open-loop) or from pseudo-open-loop measurements if operating in closed-loop.

## 2.4 Deterministic fit to a deformable-mirror

In the general case scenario – i.e. multi-conjugation – the commands are computed from $u = F\widehat{\varphi}$, where F is a deterministic least-squares volumetric fit to the DM influence functions

$$F = \left(\overline{N_{\boldsymbol{\beta}}^T N_{\boldsymbol{\beta}}} + \kappa I\right)^{-1} \overline{N_{\boldsymbol{\beta}}^T T_{\boldsymbol{\beta}}}, \tag{15}$$

where $\kappa$ (recall (1)) can be set to a often very small positive value to avoid singularities and

$$\overline{N_{\boldsymbol{\beta}}^T N_{\boldsymbol{\beta}}} = \sum_{i=1}^{n_{opt}} w_i N_{\boldsymbol{\beta}_i} N_{\boldsymbol{\beta}_i}^T \tag{16}$$

is a weighted average over $n_{opt}$ directions with relative weights $w_i$.

The DM-applied correction is in turn obtained from $\phi_{\beta_i}^{dm} = T_{\beta_i}^{dm} Nu = N_{\beta_i} u$ with u a vector of concatenated DM influence functions coefficients that form matrix N. It is a block-diagonal matrix in the multi-DM case.

We notice that the optimisation over the field is solely contained in the fitting step. We can now derive the formulations for the different AO configurations.

For ground-conjugated systems (cases of LTAO, MOAO and GLAO), $T_{\beta_i}^{dm} = I$. Further to this, a pupil-conjugated DM is isoplanatic, i.e. $N_{\boldsymbol{\beta}} = N$ and we can factor out the DM influence from (15) leading to

$$F = N^{\dagger} \overline{T_{\boldsymbol{\beta}}} = (N^T N + \kappa I)^{-1} N^T \overline{T_{\boldsymbol{\beta}}}. \tag{17}$$

The DM commands are then

$$u = N^{\dagger} W_{SA} s_{\boldsymbol{\alpha}}, \tag{18}$$

with $W_{SA} = \overline{T_{\boldsymbol{\beta}}} W_L$. This is the simplification that significantly reduces matrix size and make the whole solution more compact. Although the implicit formulation was developed with ground-conjugated systems in mind when only a single optimisation direction of interest matters (clearly MOAO, but LTAO also under certain circumstances), in practice we often seek to estimate a weighted averaged correction over a given field. As part of that development, we remark that in doing so the *implicit* formulation remains unchanged when applied to MCAO systems, as is the case of MAVIS [26] at the expense of larger matrices to evaluate the weighted sum over optimisation directions in the fitting step, i.e. the original expression from (15) expanded as

$$u = \left( \sum_{i=1}^{n_{opt}} w_i N_{\boldsymbol{\beta}_i} N_{\boldsymbol{\beta}_i}^T + \kappa I \right)^{-1} \times$$
$$\left( \sum_{i=1}^{n_{opt}} w_i N_{\boldsymbol{\beta}_i}^T \boldsymbol{\Sigma}_{\phi_{\boldsymbol{\beta}_i} \phi_{\boldsymbol{\alpha}}} \right) (\boldsymbol{\Sigma}_{ss} + \boldsymbol{\Sigma}_{\eta\eta})^{-1} s_{\boldsymbol{\alpha}}. \tag{19}$$

The trade-off between computational complexity of the fitting step in the *implicit* versus the complexity associated with the number of estimated layers in the *explicit* reconstructors is unavoidable. Quantitative estimates are provided below in Fig. 12.

## 2.5 Noise-weighted, least-squares reconstruction

A least-squares fit to the measurements is a degenerate solution leading to degraded performance [30, 50]. It has nonetheless been tested in several operational systems as the AOF [56] and GeMS [52] in what is called the virtual-DM approach – when the wavefront is not estimated independently from the DMs but exactly at the DM conjugate ranges.

In the least-squares reconstruction case, one minimises the Tikhonov functional

$$u^* = \arg \min_u \left\| s(\boldsymbol{\alpha}) - GP_{\boldsymbol{\alpha}}^{dm} u \right\|_{L_2(\Omega), \boldsymbol{\Sigma}_{\eta\eta}}^2 + \kappa \|u\|^2, \tag{20}$$

where the *de facto* fudge regularising term $\kappa$ is present to improve numerical inversion and enforce a physically meaningful reconstruction. From (20) one finds

$$u^* = \left( (GT_{\boldsymbol{\alpha}}^{dm})^T \boldsymbol{\Sigma}_{\eta\eta}^{-1} GT_{\boldsymbol{\alpha}}^{dm} + \kappa I \right)^{-1} (GT_{\boldsymbol{\alpha}}^{dm})^T \boldsymbol{\Sigma}_{\eta\eta}^{-1} s_{\boldsymbol{\alpha}}, \tag{21}$$

which does not depend on turbulence statistics whatsoever as (14) did. A closely-related tomographic reconstructor involving the so-called tomographic interaction matrix $GT_{\boldsymbol{\alpha}}^{dm}$ was tested on-sky with the MAD demonstrator at the VLT with further adaptations for stable closed-loop control [48, 62].

## 2.6 Numerical v. optical averaging for GLAO

In GLAO mode, the guide-stars are often at very large separations, ranging from 1 to tens of arcminutes. In that case the cross-covariance matrices converge to near zero and can somehow be neglected. If we set $\boldsymbol{\beta} = \boldsymbol{\alpha}$ in (19) and further neglect the noise weighing and the need to stabilise the solution via the fudge factor $\kappa$, the solution converges to

$$u = \left\langle (GN)^\dagger s(\alpha) \right\rangle_{\boldsymbol{\alpha}}, \tag{22}$$

i.e. an averaged command computed from the generalised inverse of the system interaction matrix GN in all the $\alpha$ directions. In practice the GLAO solution is the limiting case when GS separations are very large. Due to that configuration, the tomographic solution tends to the straight average of correction on each GS direction. This optically-granted averaging can be mimicked in (19) by optimising over a great number of directions spanning a large field. The latter is here called "numerical" averaging and should in principle always outperform the former if a good knowledge of the priors is held. The optical averaging lifts this need.

## 2.7 Summary of this section

This section entails important results. We have seen that the LMMSE solution

1. is found as a limiting-case of a state-space-based (linear) optimal control solution
2. can be expressed as two mathematically equivalent forms, dubbed *implicit* and *explicit*
3. is split into two steps: a stochastic wavefront estimation regardless of which AO mode is in use, followed by a deterministic fitting step independent from the wavefront and noise statistics
4. is model-based with two stages: learning or identifying the model from prior knowledge or empirically from telemetry followed by on-line exploitation in real-time
5. has been used with different names in the community, yet the underlying solution is fundamentally the same expressed in different mathematical bases

## 3 Modelling SR-enabling formulations

We have already remarked that super-resolution is native to tomographic systems on account of the naturally-occurring offsets between samples of turbulence when tracing from upper layers to the WFS subaperture intercepts along the lines of sight of two distinct guide stars. It is therefore expected that we can accommodate such offsets at the ground with minimal effort should one adopt a ground-enabling SR optical design.

As we shall see shortly, this is indeed possible both in the *explicit* and *implicit* cases. In the former case we modify the ray-tracing operator to the new subaperture (offset) coordinates whereas in the latter we evaluate the auto- and cross-covariance matrices directly on the, typically offset and/or rotated, ground-coordinates and their back-projections along the guide-stars lines-of-sight.

### 3.1 Common terms

#### 3.1.1 Auto- and cross-covariance matrices

The auto-covariance matrices $\mathbf{\Sigma}_{\varphi\varphi}$ are computed by summing over all the layers the covariance between all the samples located at the WFS intercepts, i.e.

$$\mathbf{\Sigma}_{\varphi\varphi}(\alpha_i, \alpha_j) = \sum_l \mathcal{C}(|\boldsymbol{\rho}_l(\boldsymbol{\alpha}_i) - \boldsymbol{\rho}_l(\boldsymbol{\alpha}_j)|), \qquad (23)$$

where

$$\mathcal{C}(\rho) = \left(\frac{L_0}{r_0}\right)^{5/3} \frac{\Gamma\left(\frac{11}{6}\right)}{2^{5/6}\pi^{8/3}} \left(\frac{24}{5}\Gamma\left(\frac{6}{5}\right)\right)^{5/6} \times$$
$$\left(\frac{2\pi\rho}{L_0}\right)^{5/6} K_{5/6}\left(\frac{2\pi\rho}{L_0}\right), \tag{24}$$

if $\rho > 0$ and

$$\mathcal{C}(0) = \left(\frac{L_0}{r_0}\right)^{5/3} \frac{\Gamma\left(\frac{11}{6}\right)\Gamma\left(\frac{5}{6}\right)}{2\pi^{8/3}} \left(\frac{24}{5}\Gamma\left(\frac{6}{5}\right)\right)^{5/6}, \tag{25}$$

otherwise. $L_0$ is the outer scale of turbulence, $r_0$ the Fried's parameter, $\Gamma(\cdot)$ the "gamma" function, and $K_{5/6}(\cdot)$ a modified Bessel function of the third order. Parameter $\rho = |\boldsymbol{\rho}_l(\boldsymbol{\alpha}_i) - \boldsymbol{\rho}_l(\boldsymbol{\alpha}_j)|$ is shorthand notation for a square matrix of distances between all the WFS intercepts along guide-star direction $\boldsymbol{\alpha}_i$ and $\boldsymbol{\alpha}_j$ projected at height $h_l$. An illustration of (23) is shown in Fig. 2.

The intercept coordinates are given by

$$\boldsymbol{\rho}_l(\boldsymbol{\alpha}_i) = \{\xi x_{0,i} + \alpha_{x,i}h_l; \xi y_{0,i} + \alpha_{y,i}h_l\}, \tag{26}$$

with $\xi = 1 - h_l/h_{LGS}$ the cone compression factor when the guide-stars are at a finite height, offset by $\boldsymbol{\alpha}_i h_l$ and $\{x_{0,i}, y_{0,i}\}$ are the $i-th$ WFS ground coordinates. The matrix of distances is therefore of size $n_{intercepts} \times n_{intercepts}$. Using vector programming, the distances can be easily computed by expressing the coordinates as complex numbers vectorised as a column and computing

$$|\boldsymbol{\rho}_l(\boldsymbol{\alpha}_i) - \boldsymbol{\rho}_l(\boldsymbol{\alpha}_j)| = |(x_i + iy_i) - (x_j + iy_j)^{\mathrm{T}}|. \tag{27}$$

When the WFSs are offset and rotated with respect to the nominal reference frame, the intercepts in the pupil plane can be conveniently modified to account for such features. Figure 6 depicts two instances, one offset and one rotated.

### 3.1.2 Discrete-gradient operator

The discrete-gradient operator can be modelled with over-sampling factors 1, 2 or 4, as shown next. We use a stencil representation which is akin to a convolution kernel. To enable SR, the default over-sampling is 2 for which we provide its representation below.

If we assume a WFS gradient operator with a $3 \times 3$ stencil [39] then we have

$$\mathrm{stencil}(\mathbf{G_x}) = \frac{1}{d}\begin{bmatrix} -1/4 & 0 & 1/4 \\ -1/2 & 0 & 1/2 \\ -1/4 & 0 & 1/4 \end{bmatrix}, \tag{28}$$

$$\mathrm{stencil}(\mathbf{G_y}) = \mathrm{stencil}(\mathbf{G_x})^{\mathrm{T}}. \tag{29}$$

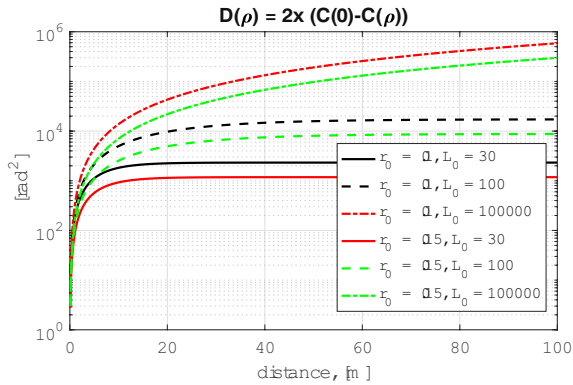

**Fig. 2** Atmospheric phase structure function (1D) computed from (23) as $\mathcal{D}(\rho) = 2(\mathcal{C}(0) - C(\rho))$. For distances above $L_0$ the function saturates

This applies to fully illuminated sub-apertures. Yet, for sub-apertures located at the inner and outer rims or under telescope struts, the illumination is only partial. We have developed a means to conserve the wavefront gradient, conveniently expanded as bilinear spline functions by including a weighting by the amplitudes at each of the 9 points. For all those cases, we have

$$\text{stencil}(\mathbf{G_x}) = \frac{1}{d} \begin{bmatrix} g_{0,0} & g_{0,1} & g_{0,2} \\ g_{1,0} & g_{1,1} & g_{1,2} \\ g_{2,0} & g_{2,1} & g_{2,2} \end{bmatrix}, \tag{30}$$

where

$$
\begin{aligned}
g_{0,0} &= -\frac{2(a_{0,0}^2 + a_{0,1}^2) + a_{1,0}^2 + a_{1,1}^2}{3(a_{0,0}^2 + a_{0,1}^2 + a_{1,0}^2 + a_{1,1}^2)} \\
g_{0,2} &= \frac{2(a_{0,1}^2 + a_{0,2}^2) + a_{1,1}^2 + a_{1,2}^2}{3(a_{0,1}^2 + a_{0,2}^2 + a_{1,1}^2 + a_{1,2}^2)} \\
g_{2,0} &= -\frac{a_{1,0}^2 + a_{1,1}^2 + 2(a_{2,0}^2 + a_{2,1}^2)}{3(a_{1,0}^2 + a_{1,1}^2 + a_{2,0}^2 + a_{2,1}^2)} \\
g_{0,1} &= -(g_{0,0} + g_{0,2}) \\
g_{2,1} &= -(g_{2,0} + g_{2,2}) \\
g_{1,0} &= -(g_{0,0} + g_{2,0} + 2) \\
g_{1,2} &= -(g_{0,2} + g_{2,2}) + 2 \\
g_{1,1} &= -(g_{1,0} + g_{1,2})
\end{aligned}
\tag{31}
$$

with $a_{i,j} \in \{0 \cdots 1\}$ is the amplitude of each of the nine discrete phase points the stencil relates to. By choosing $a_{i,j} = 1$ i.e. a fully illuminated sub-aperture, one gets the same stencil as in (28).

The key assumption behind enabling super-resolution is that the wavefront can be estimated at a higher resolution than that nominally defined by the sampling of each WFS. The $3 \times 3$ stencil enables this by allowing the wavefront to be expressed at twice the spatial sampling rate of the WFS. We could further extend the resolution using the stencil

$$
\text{stencil}(G_x) = \frac{1}{d} \begin{bmatrix} -1/16 & 0 & 0 & 0 & 1/16 \\ -3/16 & 0 & 0 & 0 & 3/16 \\ -1/2 & 0 & 0 & 0 & 1/2 \\ -3/16 & 0 & 0 & 0 & 3/16 \\ -1/16 & 0 & 0 & 0 & 1/16 \end{bmatrix} \tag{32}
$$
$$
\text{stencil}(G_y) = \text{stencil}(G_x)^{\text{T}}
$$

However, as explained in [55] estimating wavefronts at spatial frequencies beyond twice the WFS spatial cut-off will become in practice difficult: the zeros of the $sinc^2$ sensitivity function will significantly increase the noise propagation and despite appropriate regularisation, it will prove more challenging in operational scenarios. For this reason we turn our attention to estimating the wavefront with an over-sampling factor of 2$x$ only.

### 3.1.3 Noise covariance matrix

The noise between sub-apertures is considered statistically independent. For this reason, the noise covariance matrix is either diagonal (case of compact natural sources) or block diagonal (case of laser guide star elongated spots) and therefore quite sparse as well.

Incorporating noise priors that account for the sodium layer profile into the noise covariance matrix is critical for performance, especially in ELT AO systems ([11, 65]). While the improvement in wavefront error is around 10 nm RMS for an 8 m telescope, it increases to 100 nm RMS for a 40 m telescope. A noise model for sodium laser elongated – truncated and non-truncated – spots was presented in [57], representing an accurate and practical solution suitable for implementation. A complementary, alternative solution is the use of matched filtering for the spot centroid computation ([19, 36, 37]) which reduce the impact of truncation [8]. Figure 3 illustrates the sparse pattern for an elongated laser guide star case, where off-diagonal correlations show up between the 'x' and 'y' measurements on account of the rotational spot elongation, as shown in Fig. 4 for side-launch laser telescopes.

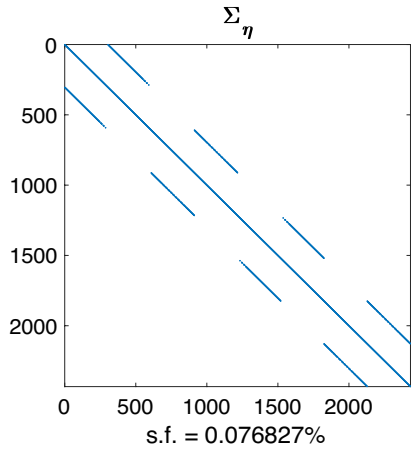

**Fig. 3** Noise covariance matrix

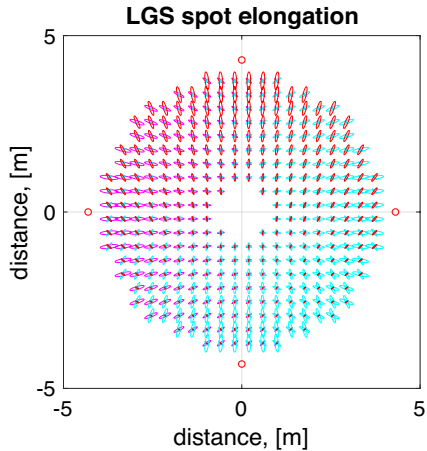

**Fig. 4** Representation of the spot elongations for a 4 laser system. The red dots represent the laser launch telescope locations and the four coloured patterns the elongation created by each

## 3.2 Sparse explicit tomography

### 3.2.1 Ray-tracing

In our formulation, the ray-tracing matrices encode the propagation from atmospheric layers to the pupil-plane. They concatenate bilinear interpolation coefficients and are therefore quite sparse, containing at most four non-zero entries per phase point per layer. If the WFSs are offset and rotated with respect to the nominal reference frame, the intercepts in the pupil plane can be conveniently modified to account for such features. The propagator $T_{\boldsymbol{\alpha}}(\boldsymbol{\Delta}_{\boldsymbol{\alpha}_i}, \theta_i)$ thus becomes a function of $(\boldsymbol{\Delta}_{\boldsymbol{\alpha}_i}, \theta_i)$, i.e. a 2D offset per WFS channel and a rotation, with its modified coordinates first rotated about the origin

$$\begin{bmatrix} x' \\ y' \end{bmatrix} = \begin{bmatrix} cos(\theta) & -sin(\theta) \\ sin(\theta) & cos(\theta) \end{bmatrix} \begin{bmatrix} x \\ y \end{bmatrix}, \tag{33}$$

and then off-set

$$\begin{bmatrix} x'' \\ y'' \end{bmatrix} = \begin{bmatrix} x' + \Delta_{\boldsymbol{\alpha}_i}^x \\ y' + \Delta_{\boldsymbol{\alpha}_i}^y \end{bmatrix}. \tag{34}$$

Using homogenous coordinates, we get

$$\begin{bmatrix} x'' \\ y'' \\ 1 \end{bmatrix} = T(\theta_i) R(\boldsymbol{\Delta}_{\boldsymbol{\alpha}_i}) \begin{bmatrix} x \\ y \\ 1 \end{bmatrix}, \tag{35}$$

with

$$\begin{bmatrix} x'' \\ y'' \\ 1 \end{bmatrix} = \begin{bmatrix} cos(\theta) & -sin(\theta) & 0 \\ sin(\theta) & cos(\theta) & 0 \\ 0 & 0 & 1 \end{bmatrix} \begin{bmatrix} 1 & 0 & \Delta_{\boldsymbol{\alpha}_i}^x \\ 0 & 1 & \Delta_{\boldsymbol{\alpha}_i}^y \\ 0 & 0 & 1 \end{bmatrix} \begin{bmatrix} x \\ y \\ 1 \end{bmatrix}. \tag{36}$$

### 3.2.2 Sparse approximation of the phase covariance functions

The covariance matrix of the layered phase distortion, or rather its inverse as per (13), i.e. $\boldsymbol{\Sigma}_{\varphi\varphi}^{-1}$, can be approximated as a sparse matrix with a discrete Laplacian operator L as $\boldsymbol{\Sigma}_{\varphi\varphi}^{-1} \approx L^T L$ [28]. The discrete Laplacian takes the form of a scaled convolution kernel characterised by the stencil

$$L = \begin{bmatrix} 0 & 1/4 & 0 \\ 1/4 & -1 & 1/4 \\ 0 & 1/4 & 0 \end{bmatrix}. \tag{37}$$

For phase points at the edge of the reconstruction grid, the coefficients are 'folded' back into the grid to ensure that their sum remains zero.

In conclusion, the matrices that compose the reconstructor–$T_{\beta}$, $T_{\alpha}$, G, and $\boldsymbol{\Sigma}_{\eta\eta}^{-1}$ –can be effectively represented using sparse approximations. Moreover, in designs incorporating offset and/or rotation for SR, the ray-tracing operation only is adapted to account for these features without altering its inherent sparsity structure. This can be visualised on Fig. 5, where, clearly, the ray-tracing encompasses all modifications, remaining with the same sparse density.

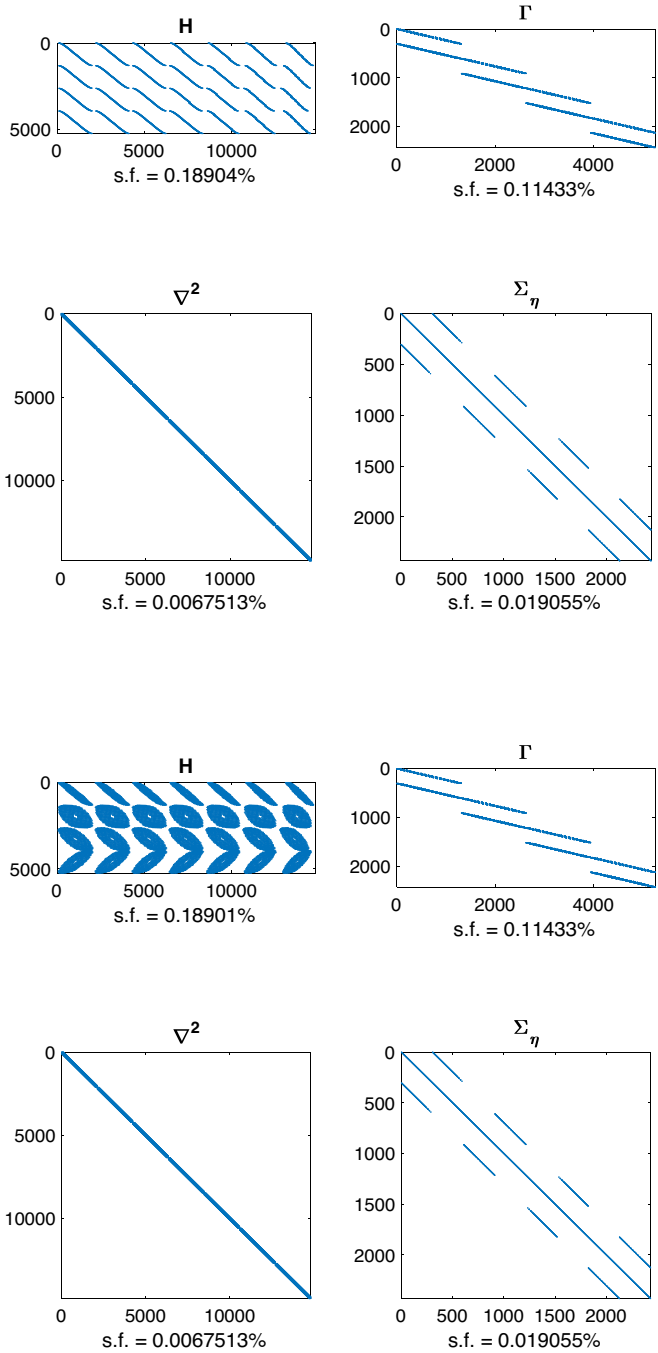

**Fig. 5** Top: Discrete, zonal, sparse matrices for the nominal case. Bottom: same for the rotated case

## 3.3 Compact, implicit tomography

In the implicit reconstruction case, the formulation no longer allows for a sparse representation as in the explicit case. However, the pupil-plane covariance matrices encapsulate all angular correlations integrated across atmospheric layers. The resulting matrix to be inverted is of the same dimension as the measurement vector s, rather than the full three-dimensional phase vector $\varphi$, leading to substantial computational savings in terms of storage, memory usage, and inversion complexity.

Figure 6 depicts cross-covariance matrices for the nominal case (all sensors aligned), versus offset and rotated cases represented in Fig. 7. Similarly to the *explicit* reconstructor, where the ray-tracing matrix T encoded these transformations, in the *implicit* reconstructor it is directly the cross-covariance matrix via the baselines

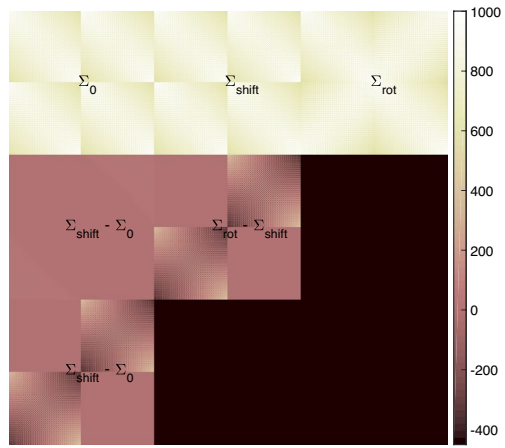

**Fig. 6** LGS cross-covariance functions for three cases: nominal ($\Sigma_0$), with shifts ($\Sigma_{shift}$) and with rotations ($\Sigma_{rot}$). The mid and lower rows show differences between those matrices as labelled

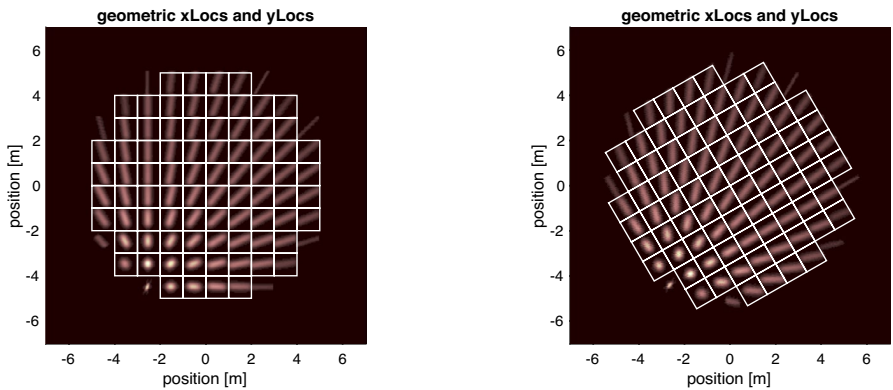

**Fig. 7** Illustrative configurations with spot elongation from a laser launch telescope placed at the bottom left. This shows that rotations could improve performance two-fold: by minimising laser spot truncation and enabling super-resolution. The impact of a nominal vs a rotated frame on the covariance matrices in shown in Fig. 6

established between all the pupil-plane grid intercepts across multiple directions that encapsulate such information. More specifically, as shown in Fig.6, the differences between the cross-covariance matrices reveal that the signatures of the transformations primarily appear in the cross-terms, i.e., the off-diagonal blocks.

### 3.4 Summary of this section

The main takeaway points from this section are the following:

1. the *implicit* and *explicit* LMMSE formulations lead to different matrix structures that can be exploited to optimise processing with custom arrangements
2. in either formulation, super-resolution can be naturally incorporated by expressing the ray-tracing or covariance baselines on distinct wavefront meshes at the pupil plane. While this results in the loss of certain matrix structures (e.g., Toeplitz), it does not affect the performance of dense-matrix reconstructors in real-time processing, which remains unchanged.

## 4 Hard real-time processing: iterative or not?

### 4.1 Solving for the wavefront using iterative implementations

Iterative implementations of both the sparse-matrix-based explicit and the implicit reconstructors have been extensively studied in the past two decades. Regarding the former, the original formulation was developed by [28] with several implementations put forward exploiting different block-structured sparse-matrix aspects: [40] using multigrid methods with preconditioning, [27] expanding to Gauss-Seidel iterations and conjugate-gradient methods, [80] using a Fourier-domain circulant-block approximation, [35] formulating the iterative solution in the spatial frequency domain, [79] generalising the method for the fitting step also, [73] using a fractal description of turbulence exploited in the regularisation step (but otherwise using a preconditioned conjugate-gradient algorithm), [68] using a finite element wavelet hybrid algorithm (FEWHA), Kaczmarz and gradient-based formulations to cite the most prominent.

Generally speaking, for the iterative case, (13) can be split into three steps:

$$\begin{cases} \text{i)} & \zeta = T_\alpha^T G^T \Sigma_{\eta\eta}^{-1} s, \\ \text{ii)} & (T_\alpha^T G^T \Sigma_{\eta\eta}^{-1} G T_\alpha + \Sigma_{\varphi\varphi}^{-1})\hat{\varphi} = \zeta, \\ \text{iii)} & \hat{\phi} = T_\beta \hat{\varphi}, \end{cases} \quad (38)$$

where *ii*) in (38) is solved for iteratively.

The *implicit* reconstructor also admits a similar implementation, under the assumption that all wavefront sensors are of the same order and perfectly aligned with respect to one another. This leads to a Toeplitz structure in the auto-covariance matrices which is exploited in the solver to estimate for the wavefront without explicit matrix inversion [20, 54]. This structure remains in place under spatial shifts yet breaks down in a more general case with rotation, different sampling and differential distortions between the wave-front sensors.

Were an iterative implementation to be deployed in real-time, the tomographic reconstruction would take the form:

$$
\begin{aligned}
&a) \quad \left(\mathbf{\Sigma}_{ss} + \mathbf{\Sigma}_{\eta\eta}\right)\zeta = \mathrm{s}, \\
&b) \quad \hat{\phi} = \mathbf{\Sigma}_{\phi s}\zeta,
\end{aligned}
\tag{39}
$$

where (39)$-a$ is solved for iteratively and (39)$-b$ requires one matrix-vector multiplication using the solution from (39)$-a$. Although the covariance matrices are not sparse, their Toeplitz structure enables efficient vector multiplication in the Fourier domain, significantly reducing computational cost through the use of fast Fourier transform (FFT) algorithms.

### 4.2 Direct matrix inversion and use of dense VMMs

Yet, dense matrix-vector multiplications are admittedly the preferred implementations for hard real time processing for its best use of memory access bandwidth, parallelisation and pipelinability features [10, 77]. Iterative methods, nonetheless, were (and are still) rather useful for physical-optics simulations and fast performance evaluation. And as it turns out, to compute the very (dense) reconstructor.

In order to do that, the reconstructor is assembled by applying the reconstructor and DM fitting steps on the identity matrix [77], in what is otherwise a rather standard matrix inversion technique [41]. For the *explicit* reconstructor one can solve for the $i - th$ reconstructor column by equating

$$
\mathrm{R}[:, i] = A_{\mathrm{L}}^{-1} b_{\mathrm{L}} \mathrm{I}[:, i],
\tag{40}
$$

where $A_{\mathrm{L}}^{-1}$ and $b_{\mathrm{L}}$ are found in (38).

The *implicit case* follows the same principle – that we omit here for the sake of conciseness – using the iterative form of (39) instead.

### 4.3 Real-world cases with dense vector-matrix multiplies

This covariance-based reconstructor has already been used with on-sky tomographic MOAO systems on current large telescopes: The spatial-angular reconstructor for RAVEN on the 8 m Subaru Telescope [17] and the *Learn & Apply* algorithm for CANARY on the 4 m William Herschel Telescope [74]. These reconstructors use synthetic covariance matrices based on a prior information of atmospheric parameters calibrated with on-sky WFS measurements. The synthetic covariance matrices and reconstructors are updated somewhere from 10 seconds to a few minutes depending on the atmospheric conditions, bandwidth and special geometric considerations. In the *Learn & Apply* algorithm, the reconstructor takes into account not only atmospheric conditions but also system mis-registration manifesting as optical alignment errors (pupil shifts, rotations, and magnifications).

In the tomographic AO systems so far developed for the current 8 m-class telescopes, the complexity of the system i.e. the number of measurements is $1\text{-}3 \times 10^3$ (e.g. GeMS with 5 LGSs and $20 \times 20$ SH-WFSs; KAPA with 4 LGSs and $20 \times 20$ SH-WFSs; RAVEN with 3 NGSs + 1 LGS with $10 \times 10$ SH-WFSs). Therefore, the tomographic reconstructor is computed and updated within a fraction of the time a reconstructor takes to become stale, something which is system-specific and can range from tens of seconds to tens of minutes [34] and the wavefront reconstruction is done with one matrix-vector multiplication like (18).

In addition to systems in 10m-class telescopes such as the AOF [56], KAPA [78], MAVIS [5], GNAO [70], Ultimate-Subaru [71], next generation GSMT's will also adopt dense-matrix implementations. Cases of the TMT's NFIRAOS [76], ELT's Harmoni [44] and Morfeo [3].

### 4.4 Summary

The main points of this section are:

1. Dense-matrix, non-iterative formulations are preferred for real-time processing, provided the effectiveness of their implementation making more optimal use of memory bandwidth,
2. iterative approaches are still of use to invert the very matrices to be used under the previous point,
3. both *implicit* and *explicit* formulations can be mapped into an iterative solver,
4. super-resolution has been or will soon be adopted on all tomographic systems on 8 m- and 40 m-class telescopes.

## 5 Sample numerical examples on [8-10]m-class telescopes

In this section we provide sample results for new-generation tomographic AO systems on [8-10]-m facility-class telescopes implementing LTAO and MCAO

1. GNAO on the Gemini north telescope, featuring 4 lasers on a 5-10" (LTAO) and 40-50" (GLAO) radius feeding a 20x20 SH-WFS AO system [70]. Its first client instrument, GIRMOS, an Infrared Multi-Object Spectrograph will provide science from $1\mu m$ upwards [16]
2. Ultimate Subaru's narrow and wide field modes, namely the ULTIMATE-START with an LTAO behind AO-188 assisted by 4 LGSs on a 10-40" radius sampled with 32x32 SH-WFSs and driving a 3228 DM [71] will provide AO correction in the wide wavelength range between 600 nm and 2500 nm [2]
3. MAVIS [5, 6] on the VLT with 8x 40x40 SH-WFS, each with 6x6 pixels/subaperture providing increased performance at shorter wavelengths towards the visible.

All these systems adopted the *implicit* reconstructor formulation. Moreover, provided the *small* system sizes, the matrix inversion is made using standard linear algebra techniques without recurring to the custom iterative formulations shown in §4. At the time of writing, the latter seem reserved to ELT-sized systems. At this stage only NFIRAOS on the TMT [76] advocates the use of the combination of the *explicit* reconstructor obtained via iterative, row-wise inversion.

## 5.1 Laser-tomography systems

To showcase the impact of super-resolution we choose a telescope with a primary mirror diameter $D = 8$ m with the atmospheric conditions tabulated in Table 1.

We have considered three different observing conditions at Mauna Kea, $r_0$ [75p]= 13.5 cm, $r_0$[median]= 18.6 cm and $r_0$[25p]= 24.7cm (values within squared brackets are the statistical percentiles). Table 2 provides the different system configurations used in the simulations. They have as common parameters: $RON = 0.5\,e-$, excess noise $= 1.41\,e-$, $QE = 90\%$, pixel size = 0.8", 6x6 pixel per subaperture on the LGS WFS. For the DMs, we used Gaussian functions with a mechanical coupling of $25\%$. The LGS beacons provide a return flux $= 43.1\,ph/s/m^2/W$ with a $FWHM_{on-sky} = 1.5$".

To understand the effect of a higher order DM on the performance, we evaluate the gain from a DM:$20 \times 20 \rightarrow$DM:$30 \times 30$ in the $J-$ and the $V-$bands (see Fig. 8). Only the fitting error is considered here, computed using

**Table 1** Atmospheric conditions used during the simulations

| Elevation (km)      | 0    | 0.5  | 1    | 2    | 4     | 8    | 16   |
|---------------------|------|------|------|------|-------|------|------|
| Wind speed(m/s)     | 5.6  | 5.7  | 6.5  | 7.5  | 13.3  | 19.0 | 12.1 |
| Wind direction (°)  | 190  | 255  | 270  | 350  | 17    | 29   | 66   |
| Turbulence fraction | 0.45 | 0.12 | 0.04 | 0.05 | 0.11  | 0.09 | 0.11 |

**Table 2** System configurations

| Case              | LGS SH WFS     | DM             | #LGS |
|-------------------|----------------|----------------|------|
| No super-resolution | $20 \times 20$ | $21 \times 21$ | 4    |
| Super-resolution    | $20 \times 20$ | $30 \times 30$ | 4    |
| Ideal               | $30 \times 30$ | $31 \times 31$ | 4    |

**Fig. 8** Theoretical gain of Strehl-ratio (fitting-error only) in $J-$band (1250 nm, dashed) and $V-$band (650 nm, solid). The super-resolution case considers a DM$_{30 \times 30}$ and the nominal, no super resolution is DM$_{20 \times 20}$

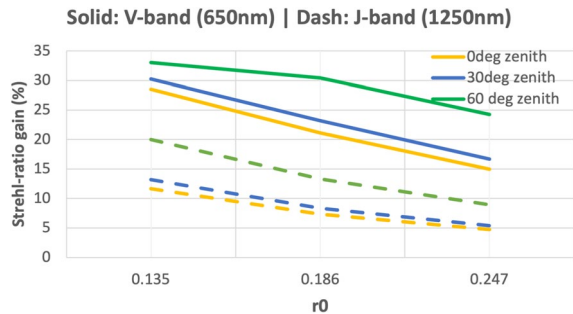

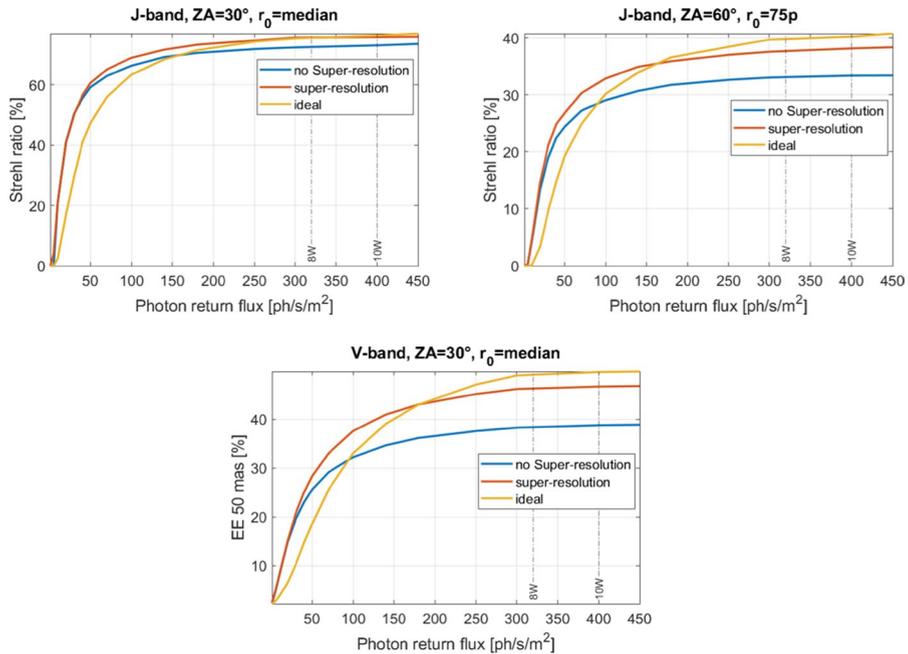

**Fig. 9** Performance in Strehl ratio versus the photon flux return. left: the science is in *J*-band, ZA= $30°$, $r_0$ =median. Middle: *J*-band, ZA= $60°$, $r_0$ =75p. Right: V-band EE$_{50mas}$. We also represent on two vertical lines the nominal power of an individual laser at 10W and 20W

$$\sigma^2 = \alpha \left( \frac{D/n_{actuators}}{r_0} \right)^{\frac{5}{3}}, \tag{41}$$

where we assume $\alpha = 0.23$ for a regular grid, squared geometry actuator layout.

In Fig. 8, we show how super-resolution improves performance as a function of $r_0$. The expected performance boost is larger for shorter wavelengths and poorer observing conditions.

To fully assess the AO performance, we carried out E2E simulations using +OOMAO+ [14]. We consider a global throughput of 0.3 (before LGS WFS camera) and three cases: with SR, without SR and an ideal case where the large number of DM actuators is matched by as many subapertures.

First we evaluate the performance in the *J*-band at ZA= $30°$ in the median $r_0$ conditions.

In Fig. 9-left, as anticipated, the performance of the "Ideal" case with more subapertures drops before the two other cases. The super-resolution case performs better than the one without super-resolution and does not propagate more error.

In Fig. 9-middle, the relative performance gain with super-resolution is higher than in the previous case as the fitting error increases under these atmospheric conditions. Here super-resolution not only improves substantially the performance, it gives extra robustness to the LGS photon return flux variations.

In Fig. 9-right, we look at the EE in 50 mas (radius) in the *V*-band at ZA= 30° and median $r_0$. Again, the super-resolution case increases substantially the performance when the laser power is more limited and the LGSs relatively fainter.

## 5.2 Multi-conjugate AO systems: The case of MAVIS

The MCAO Assisted Visible Imager and Spectrograph (MAVIS) [75] targets visible wide-field correction on the VLT, utilising eight laser guide stars, three natural guide stars, and three deformable mirrors (including the deformable secondary mirror of the UT4). The modest-sounding goal of 15% Strehl-ratio becomes a real challenge when applied at 550 nm wavelength, over a 30 arcsec diameter disk, for at least 50% of the sky at the galactic poles.

MAVIS has an incredibly tight error budget, so super-resolution was quickly incorporated into the MAVIS design upon its dissemination into the AO community (in particular, since [58]). Upon investigating the reconfiguration of the MAVIS LGS WFSs to harness SR, it was immediately clear that the precise choice of SR-enabling geometry was much less important than simply avoiding the homologous case (when all wavefront sensors sample the pupil identically). That is, the benefit of super-resolution is relatively flat near the optimal configuration (see, e.g., [13, 15]). One cannot do worse than aligning each WFS to sample the pupil with the same geometry, and avoiding this pathological default case will immediately earn most of the benefits of super-resolution.

In MAVIS, the implementation of super-resolution is achieved by *clocking* the WFSs with respect to each other, so that the footprint of the pupil on the WFS micro-lens array (MLA) is rotated for each WFS. Additionally, we loosened the tolerance for transverse alignment of the MLA, allowing any "accidental super-resolution" and simplifying the alignment procedure. The benefit of SR in MAVIS can be seen by comparing the performance in end-to-end simulations (using COMPASS [33]). For the nominal MAVIS configuration (with WFS clocking), we performed a simulated exposure of 5 seconds (5000 frames) and record the average long exposure Strehl ratio over the science FoV. This is repeated for a varying number of tomographic optimisation targets (to improve the performance uniformity across the field). Finally, the same experiments are run without WFS clocking (homologous pupil sampling). The results are summarised in Fig. 10.

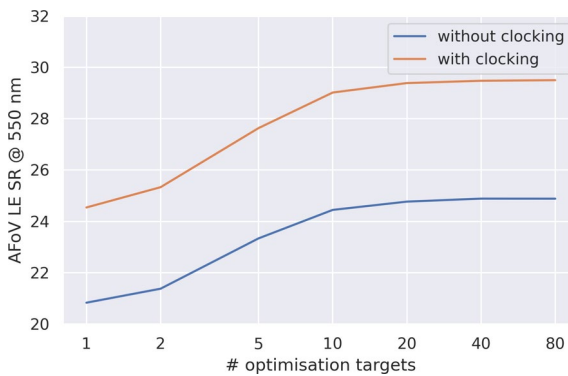

**Fig. 10** Comparison of performance in MAVIS with and without WFS clocking (i.e., super-resolution)

There are two main points to note from these results:

1. Increasing the number of targets used in tomographic optimisation beyond ≈15 does not significantly improve performance.
2. Super-resolution enabling clocking of the WFSs provides an improvement of approximately 4% strehl ratio over the science FoV, which in this case equates to approximately 35 nm of wavefront error.

Note that for these simulations, we use a slope-based Predictive *Learn & Apply* reconstructor (see, e.g., [81]).

## 6 Tomography on ELT-sized telescopes

We now turn our attention to ELT-sized systems. A full-scale comparison of actual implementations and optimisations is out of scope here, but we draw inspiration from such works as [76] for the *Thirty-Meter Telescope* where a performance comparison and trade-off between the factor, number and location of over-sampled estimated wavefronts is presented for the *explicit* 6-layer reconstructor.

We limit ourselves to a subset of comparative analyses conducted for laser-tomography assisted instrument on the ELT Harmoni and the multi-conjugate AO platform MORFEO. More complete information can be found in [44] and [3] respectively. Each is covered in a subsection below.

In either of these systems, the ELT quaternary mirror (M4) will be used as wavefront corrector. In Fig. 11 we depict an illustration of the diffraction-limited PSF over-plotted on the M4 fitting error – showing the M4 correction bounds. These turn out to be hexagonally-shaped on account of the M4 haxogonal layout.

Key to our discussion is the computational complexity of SR-enabling reconstructors. This, as we have argued, is largely offset to the soft-real-time when the models are populated and inverted. The hard-real-time does not suffer substantial differences for any given system with fixed dimensions.

**Fig. 11** Illustration of M4 spatial-frequency correction bounds in comparison to the sensing limits of a 68×68 system with no super-resolution and with a 1.5x factor super-resolution which grants access to all M4-correctable frequencies

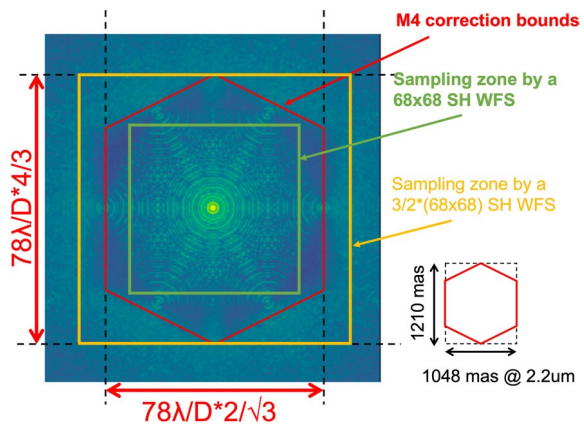

**Fig. 12** Comparison of matrix sizes for the *explicit* as a function of the number of estimated layers (red) and the *implicit* reconstructor as a function of the number of guide-stars (black)

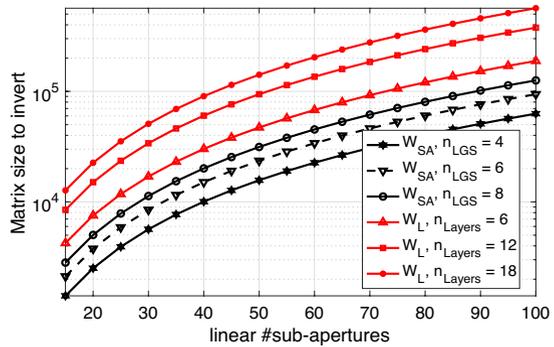

In Fig. 12 we show the evolution and comparison of matrix sizes to be inverted off-line in the *explicit* and *implicit* cases. The former largely depends on the number of layers to be estimated whereas the latter depends on the number of optimisation directions to be evaluated. In all, for roughly equivalent systems, the *implicit* matrices to be inverted are of smaller size (by factors $5 - 10\times$). We note that [76] provides an extremely useful if not required means to invert the reconstructor by use of the very iterative algorithms studied for real-time deployment. Yet in that case the off-line problem is equated to the identity by solving as many independent problems as there are columns in the reconstructor. Should it be required, the same solution can be used in the *implicit* reconstructor. In our implementation standard linear algebra routines were used therefore discarding the need to rely on more sophisticated inversion methods, holding a more general solution.

### 6.1 Harmoni LTAO

In this section, to focus on the tomographic reconstructor and the super-resolution capability associated to it, we consider a simplified model of the Harmoni LTAO system [51]. The simulation parameters are provided in Table 3.

The performance achieved with two implementations: *implicit* vs. *explicit* with SR enabled (dubbed GSR for Geometric SR) vs disabled is shown in Fig. 13 and in Fig. 14 The SR gains translate into 36 nm rms in the *implicit reconstruction* case using dense vector-matrix-multiplies and 50 nm rms in the *explicit reconstruction* recurring to an iterative sparse matrix implementation. Further to this, the *explicit* implementation is superior to the implicit by roughly 35 nm rms in either the SR and nominal modes. It is worth mentioning that the inversion of the *implicit* reconstructor requires some fine tuning that depends mostly on the Cn2 profile and whether super-resolution is enabled or not. The optimal tuning of this inversion is out of the scope of this paper and explains the small differences between the *implicit* and *explicit* implementation presented here.

**Table 3** System configurations considered for the HARMONI LTAO scale simulation. Simulation runs for 2s at 500Hz frame-rate. Performance quoted at the centre of the field

| Telescope | Diameter | 39 m |
|---|---|---|
| | AO Pupil | ELT pupil (no spiders) |
| | Imager Pupil | Circular |
| Atmosphere | Layers | 9 from 35 layers profile |
| | | (Cerro Armazones) |
| | $r_0$@500 nm | 13.85 cm |
| | $L_0$ | 25 m |
| | Zenith Angle | 30 deg |
| LGS | Number | 6 |
| | Radius | 34 arcsec |
| | Na Profile | NA (Geometric HO-WFS) |
| HO-WFS | n | 6 |
| | nSubapertures | 68 |
| | Type | Geometric WFS |
| | GSR offsets: | |
| | Shift X [subap] | [0.35 -0.85 -0.81 0.83 -0.54 0.84] |
| | Shift Y [subap] | [-0.24 0.87 -0.26 0.26 0.27 -0.01] |
| LO-WFS | Tip/Tilt/Focus | from HO-WFS |

**Fig. 13** Histograms from H-LTAO E2E simulation with nominal median conditions (35-layer profile for Cerro Armazones), with 6x 68x68 SH-WFS subapertures controlling the ELT M4 [7]

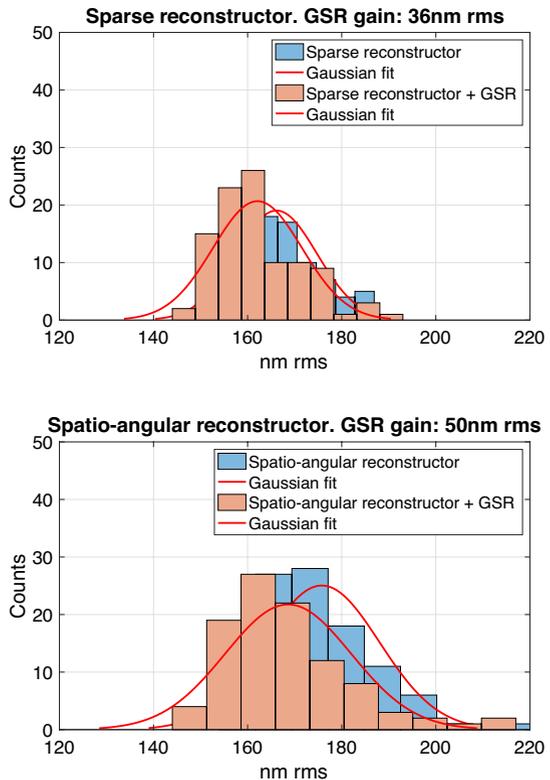

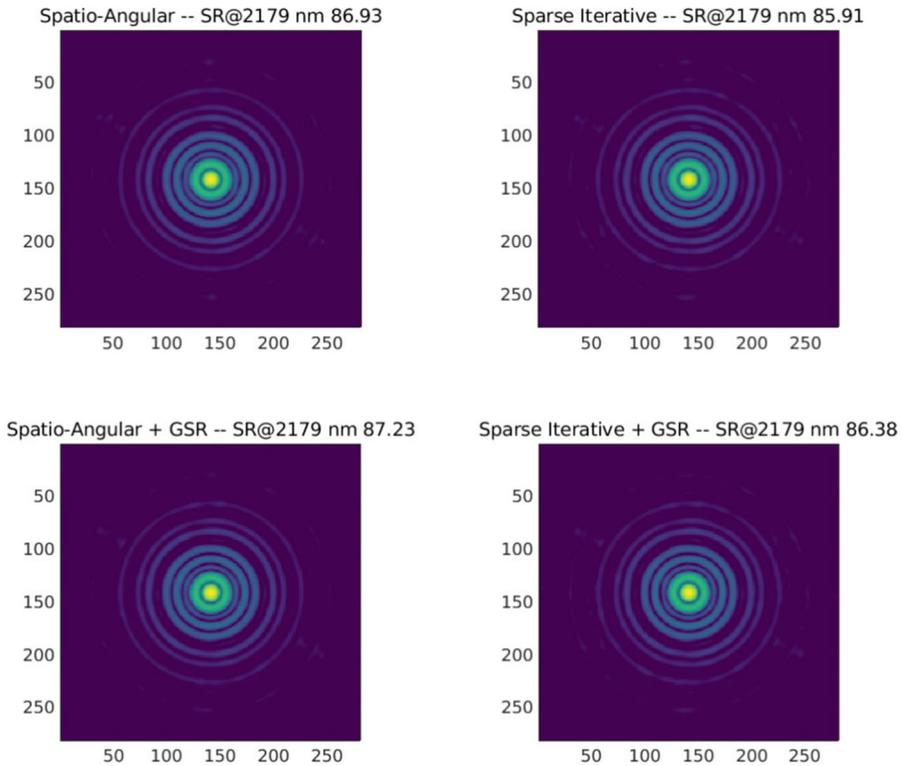

**Fig. 14** PSFs obtained for one sample case from the ones used in Fig. 13

## 6.2 MORFEO MCAO

We consider here MORFEO the MCAO system of ELT [12]. This system is designed to exploit super resolution and the number of sub-apertures of its LGS WFS is $68 \times 68$, which is a factor $\sim 0.75$ of the number of actuators of M4 seen in the pupil. The choice of this design is described in [31]. The system geometry provides free super resolution on M4 because this DM is conjugated at $\sim 600$ m above ground and rotation between LGS WFSs and the pupil – selected to align the most elongated spots on the diagonal of the sub-apertures – plus the X-Y alignment errors (identified on-line see [4]) of $\sim 0.1$ SA between the LGS WFSs and the pupil provide super resolution also on the ground layer. The tomographic reconstruction is performed modally on 10 layers and then projected onto the 3 DMs of the system. A virtual DM with a specific KL modal basis is associated with each layer, and the pitch of the virtual DM is such that the maximum spatial frequency is 50% greater than that of the nearest DM. We chose a factor of 1.5 on the maximum spatial frequencies because we need to balance the improvements provided by super-resolution with the resources required to compute the reconstructor matrix (note that this matrix must be updated during an observation to follow changes in system parameters, such as the telescope zenith angle). We run a series of simulations to show the effect of super-resolution, considering 4 different cases:

- No super resolution - no rotations are introduced between the LGS WFSs and the virtual DMs has a minimum pitch equal to 0.57 m, which is the sampling of the LGS WFSs
- Free super resolution - no rotations are introduced between the LGS WFSs and the pupil and the virtual DMs has a minimum pitch equal to 0.38 m
- Super resolution - rotations are introduced between the LGS WFSs and the pupil and the virtual DMs has a minimum pitch equal to 0.38 m
- High sampling without super resolution - the LGS WFSs have $80 \times 80$ sub-apertures, no rotations are introduced between the LGS WFSs and the pupil and the virtual DMs has a minimum pitch equal to 0.38 m

We consider the median atmospheric conditions of ELT, with a seeing of 0.65 arcsec, a zenith angle of 30deg, high flux for both LGSs and NGSs and we did not introduce spot elongation due to sodium profile thickness to focus on the super resolution (all parameters are reported in Table 4). The SR are the raw ones and did not include the full error budget of the system (i.e. we did not add NCPA error, manufacturing tolerances, ...). The results in term of K band SR are reported in Table 5 together with the differences in term of wavefront error with respect to the first case. The largest improvement is obtained just by letting the system reconstruct an higher number of modes and exploiting the super resolution provided by the different line of sight of the system. Adding rotation and shifts provide a small improvement because increasing the maximum sensing spatial frequency on the ground layer can not be exploited by a DM correction, because as written above M4 is not conjugated in the pupil. The difference with the case with the sampling equal to the one of M4 is small compared to the overall raw error ( 230 nm RMS), just 49.5 nm RMS.

**Table 4** System configuration considered for the MORFEO scale simulations. The integration time is 2 s and the frame-rate is 500 Hz

| Telescope | Diameter | 39 m |
|---|---|---|
| | Pupil | ELT pupil with spiders |
| Atmosphere | Layers | 35 layers profile |
| | | (Cerro Armazones) |
| | $r_0$@500nm (z=0deg) | 15.5 cm |
| | $L_0$ | 25 m |
| | Zenith Angle | 30 deg |
| LGS | Number | 6 |
| | Radius | 45 arcsec |
| | Na Profile | NA |
| HO-WFS | n | 6 |
| | nSubapertures | 68 |
| | Type | diffractive WFS |
| | GSR offsets: | |
| | rotations [deg] | [6.2, 14.2, -6.2, 6.2, 14.2, -6.2] |
| LO-WFS | n | 3 |
| | nSubapertures | 2 |
| | Type | diffractive WFS |
| DM | n | 3 |
| | conj. alt. [m] | [600, 6500, 17500] |
| | pitch [m] | [0.5, 1.2, 1.4] |

**Table 5** Summary of the results of MORFEO simulations. The second column shows the K band SR for bright GSs without sodium spot elongations, NCPA or implementation errors. The third column shows the difference in wavefront error in nm RMS compared to the case without super resolution. The Strehl ratio is long-exposure of 2 s averaged over 20″ diameter FoV

| case | SR(K) | diff. [nm RMS] |
|------|-------|----------------|
| No SR | 0.625 | 0 |
| Free SR | 0.649 | -67.8 |
| SR | 0.655 | -76.4 |
| No SR, high sampling | 0.662 | -83.9 |

## 7 Conclusion

This paper reviews and categorises the main developments made over the last three decades towards real-time atmospheric tomography arising in astronomical adaptive optics. Such developments originate chiefly from two areas of expertise: optimal control theory and computerised tomography as applied within the framework of inverse problem theory. We recall how the former required simplifications to cope with large numbers of degrees-of-freedom and how the latter required adaptations to the closed-loop negative feedback operation. Practical solutions are found somewhat in the middle.

In this paper we show how LMMSE tomography formulations, in either its three-dimensional, *explicit* layer reconstruction or its compact, spatio-angular *implicit* version fall under a very generic and widely used *Model & Deploy framework* and specially how they can accommodate **super-resolution**. We show that this leads to very minimal computational demand increase (if at all) and strictly speaking no extra hard-real-time complexity. Although this may seem somehow surprising, in reality it has to do with the number of latent degrees-of-freedom computed by the off-line models, which do not vary considerably in either case – and could even be reduced if one were to drive a given DM with fewer sensor measurements, a possibility offered by SR.

We review implementations prone to iterative implementations (using sparse matrix methods, block-Toeplitz with Toeplitz-blocks, etc) suitable for the different formulations, yet focus on the direct, dense format, vector-matrix real-time processing. This is so because the advent of massive parallel and high-performance computing has made iterative or approximate solutions largely obsolete for atmospheric tomography. It is well accepted today that regardless of how the forward model is for-mulated and the reconstructor computed (for instance using the very same iterative algorithms proposed for real-time) the implementation will recur to dense matrix-vector multiplies.

We showcase the implementation of the *implicit* reconstructor on 10m-class telescopes featuring LTAO (GNAO) and MCAO (MAVIS) tomographic systems and show likewise the performance results of SR-enabled formulations on ELT LTAO (Harmoni) and MCAO (MORFEO) instruments. We show that the *implicit* reconstructor can be inverted with only standard linear algebra techniques in both 10m and ELT systems (i.e. no custom implementation is needed, making for a more general solution) and that SR-enabled reconstruction leads to 30-60 nm rms improvement over their nominal non-SR counterparts.

The examples provided highlight the widespread adoption of super-resolution techniques in the design of contemporary tomographic systems, underscoring its critical role in enhancing imaging capabilities.

Looking further ahead, we invite our community to help in this endeavour. For instance, super-resolution can adopt complementary forms, such has including a temporal dimension (e.g. [53]) to further improve performance.

Future research could focus on investigating the optimal combination of NGSs and LGSs to achieve a desired sky coverage while reducing the required number of measurements and thus the need for bright stars. Similarly, the generalisation of SR using non-homogeneous sampling seems to us a very promising possibility, despite the currently unfeasible asymmetric and non-redundant AO designs.

**Acknowledgements** CMC acknowledges support funding with DOI 2022.01293.CEECIND/CP1733/CT0012 (doi) from the Portuguese *Fundação para a Ciência e a Tecnologia*. CTH acknowledges the computing facilities operated by Centre de données Astrophysiques de Marseille (CeSAM) data center at Laboratoire Astrophysique de Marseille (LAM) where part of this research was conducted.

**Funding** Open access funding provided by FCT|FCCN (b-on).

**Data Availability** No datasets were generated or analysed during the current study.

## Declarations

**Competing interests** The authors declare no competing interests

## Authors and Affiliations

**Carlos M. Correia[1,2] · Pierre Jouve[3,4] · Jesse Cranney[5] · Guido Agapito[6] · Cédric Taïssir Heritier[7]**

✉ Carlos M. Correia
  carlos.correia@fe.up.pt

  Pierre Jouve
  pierre.jouve@lam.fr

  Jesse Cranney
  Jesse.Cranney@anu.edu.au

  Guido Agapito
  guido.agapito@inaf.it

  Cédric Taïssir Heritier
  cedric.heritier-salama@onera.fr

[1]  Faculdade de Engenharia da Universidade do Porto, Rua Dr. Roberto Frias s/n, Porto 4200-465, Portugal

[2]  Center for Astrophysics and Gravitation, Instituto Superior Técnico, Av. Rovisco Pais 1, Lisbon 1049-001, Portugal

[3]  Space ODT, Lda, 4050 Porto, Portugal

[4]  Aix Marseille Univ, CNRS, CNES, LAM, Marseille, France

[5]  Advanced Instrumentation Technology Centre, Australian National University, Canberra, Australia

[6]  INAF - Osservatorio Astrofisico di Arcetri, Largo Enrico Fermi, 5, Firenze 50125, Italia

[7]  DOTA, ONERA, Salon F-13661, France